\newif{\ifjournal}
  \journalfalse

\ifjournal
  \documentclass[11pt]{elsart}
  \usepackage{amssymb,graphicx,natbib,txfonts}
\else
  \documentclass{paper}
\fi

\begin{document}

\ifjournal\begin{frontmatter}\fi

\title{Evolution of dark-matter haloes in a variety of dark-energy
  cosmologies}
\ifjournal
  \author[1]{Matthias Bartelmann},
  \author[2]{Klaus Dolag},
  \author[3]{Francesca Perrotta},
  \author[3]{Carlo Baccigalupi},
  \author[4]{Lauro Moscardini},
  \author[1]{Massimo Meneghetti},
  \author[2]{Giuseppe Tormen}
  \address[1]{ITA, Universit\"at Heidelberg, Germany}
  \address[2]{Dipartimento di Astronomia, Universit\`a di Padova, Italy}
  \address[3]{SISSA, Trieste, Italy}
  \address[4]{Dipartimento di Astronomia, Universit\`a di Bologna,
    Italy}
\else
  \author{Matthias Bartelmann$^1$, Klaus Dolag$^2$, Francesca
    Perrotta$^3$, Carlo Baccigalupi$^3$, Lauro Moscardini$^4$,\\
    Massimo Meneghetti$^1$, Giuseppe Tormen$^2$\\
    $^1$ ITA, Universit\"at Heidelberg, Germany\\
    $^2$ Dipartimento di Astronomia, Universit\`a di Padova, Italy\\
    $^3$ SISSA, Trieste, Italy\\
    $^4$ Dipartimento di Astronomia, Universit\`a di Bologna, Italy}
  \date{\emph{Proceedings contribution for ``Dark Matter/Dark Energy
    2004''}}
\fi

\begin{abstract}

High-resolution, numerical simulations of 17 cluster-sized dark-matter
haloes in eight different cosmologies with and without dynamical dark
energy confirm the picture that core halo densities are imprinted
early during their formation by the mean cosmological density. Quite
independent of cosmology, halo concentrations have a log-normal
distribution with a scatter of $\sim0.2$ about the mean. We propose a
simple scaling relation for halo concentrations in dark-energy
cosmologies.

\end{abstract}

\ifjournal\end{frontmatter}\else\maketitle\fi

\section{Introduction}

Having to accept that the expansion of the Universe is accelerating
today and that only $\sim30\%$ of its content is contributed by
matter, we need to search for what may be driving the accelerated
expansion. Friedmann's equations require the dominant form of matter
to have a pressure $p<-\rho c^2/3$, where $\rho$ is its density and
$c$ is the speed of light. The cosmological constant has $p=-\rho
c^2$. Generalising this, the equation of state is modified to $p=w\rho
c^2$, with $w<-1/3$. In the simplest of these models, $w$ is constant,
but it is more natural to assume that $w$ is a function of time, scale
factor or redshift. One possible, admittedly hypothetical form of
matter with such an equation of state is a self-interacting scalar
field with an interaction potential which is sufficiently larger than
its kinetic energy \citep[e.g.~][]{WE88.1,RA88.1,PE02.2}

Replacing the cosmological constant by such a hypothetical ``dark
energy'' has consequences for structure growth and the properties of
dark-matter haloes \citep{BA02.1,WE03.1,KL03.1}. We report here on our
studies of how halo concentrations change in a variety of dark-energy
models \citep{DO03.2}. This leads us to suggest a remarkably simple
scaling of halo concentrations with the linear growth factor in
dark-energy models. We indicate consequences for strong lensing by
galaxy clusters, which offer one possibility for constraining
dark-energy models. Throughout, we use present-day matter and
dark-energy density parameters of $\Omega_\mathrm{m0}=0.3$ and
$\Omega_\mathrm{Q0}=0.7$, and a Hubble constant of $h=0.7$ in units of
$100\,\mathrm{km\,s^{-1}\,Mpc^{-1}}$.

\section{Dark-Energy Models}

The continuity equation requires that the density of dark energy
changes with the scale factor $a$ as
\begin{equation}
  \Omega_\mathrm{Q}(a)=\Omega_\mathrm{Q0}\exp\left\{
  -3\int_a^1[1+w(a')]\d\ln a'
  \right\}\;.
\label{eq:01}
\end{equation}
This term replaces the usual cosmological-constant term in Friedmann's
equation. For $w=\mathrm{const.}=-1$, the cosmological-constant
behaviour is retained. For $w=\mathrm{const.}$, the scale-factor
dependence simplifies to
\begin{equation}
  \Omega_\mathrm{Q}(a)=\Omega_\mathrm{Q0}a^{-3(1+w)}\;.
\label{eq:02}
\end{equation}
If $w=-1/3$, the model behaves like an open CDM model without
cosmological constant because then $\Omega_\mathrm{Q}$ mimics the
curvature term in Friedmann's equation.

We use one model with $w=-0.6=\mathrm{const.}$ for reference, and two
models with time-varying $w$. These are the Ratra-Peebles model
\citep{RA88.1} in which the scalar-field potential is a power law, and
the SUGRA model \citep{BR00.2} which has an additional exponential
factor in the potential. Both are normalised such that $w_0=-0.83$ at
$a=1$. While $w$ is almost constant for the Ratra-Peebles model in the
relevant redshift range, it increases from $-0.83$ to $\sim-0.4$
between redshifts 0 and 2 in the SUGRA model.

If normalised to its amplitude today, structure grows earlier in
dark-energy compared to cosmological constant models. Since numerical
simulations demonstrate that dark-matter haloes keep a memory in their
cores of the mean cosmic density at their formation times
\citep{NA97.1}, haloes forming earlier are expected to have higher
core densities. Thus, at fixed mass and redshift, haloes are expected
to be more concentrated in dark-energy than in cosmological-constant
models. This was expected from analytic considerations. The work
reported here aims at testing this expectation with numerical
simulations.

\section{Numerical Simulations}

Using the Gadget code \citep{SP01.1}, we ran a large-scale
cosmological simulation of the $\Lambda$CDM model, identified massive
haloes within it, identified their Lagrangian volumes at the initial
redshift, added small-scale power and re-simulated them at much
increased resolution \citep{TO97.2}. The particle mass is
$5\times10^9\,h^{-1}\,M_\odot$. We thus created a sample of 17
clusters with final masses between $3\times10^{14}$ and
$2\times10^{15}\,h^{-1}\,M_\odot$. The $\Lambda$CDM simulation was
normalised to $\sigma_8=0.9$.

We then re-simulated this same cluster sample in dark-energy
cosmologies. For doing so, we shifted the initial redshift of the
simulation to higher values such that the earlier structure growth in
the dark-energy models was compensated. For achieving the same
density-fluctuation normalisation today, the initial redshift
$z_\mathrm{ini}$ of the dark-energy simulations needs to satisfy
\begin{equation}
  \frac{D_+(z_\mathrm{ini})}{D_+(0)}=
  \frac{D_{+,\Lambda\mathrm{CDM}}(z^\mathrm{ini}_{\Lambda\mathrm{CDM}})}
  {D_{+,\Lambda\mathrm{CDM}}(0)}\;,
\label{eq:03}
\end{equation}
where $D_+(z)$ is the linear growth factor as a function of redshift,
and $z^\mathrm{ini}_{\Lambda\mathrm{CDM}}$ is the initial redshift of
the $\Lambda$CDM simulation. It is important to also rescale the
initial velocities of the simulation particles to the higher initial
redshift.

The normalisation of the models is an open issue. The earlier
structure growth in dark-energy models increases the Integrated
Sachs-Wolfe effect and thus the amplitude of large-scale secondary
fluctuations in the CMB. A smaller fraction of the observed
fluctuations can then be attributed to the primordial CMB, thus the
normalisation of the power spectrum should be lowered.

However, we are interested in the properties of dark-matter haloes
whose linear scale is much smaller than that of the Sachs-Wolfe tail
in the CMB power spectrum. How the large-scale amplitude measured by
on CMB translates to small scales depends sensitively on the
large-scale slope of the density power spectrum. We argue that weak
gravitational lensing directly measures the power-spectrum amplitude
at the scales relevant here and should thus develop into the prime
method for normalising the power spectrum at small scales. For now, we
study two sets of normalisations. One has constant $\sigma_8=0.9$ for
all cosmological models, the other has reduced $\sigma_8$ such as to
take the enhanced ISW effect into account.

In total, we simulate our sample of 17 clusters in eight different
cosmologies: $\Lambda$CDM, Ratra-Peebles, SUGRA, and $w=-0.6$, the
latter three with two different normalisations each, and an open-CDM
model with $\Omega_\Lambda=0$ for comparison.

\section{Results}

We fit the NFW density profile to all clusters and obtained
concentration parameters for all of them at 50 output redshifts
between $z=3$ and today. For all these cluster snapshots, we also
compute analytically expected concentrations according to the
algorithms proposed by \cite{NA97.1,BU01.1,EK01.1}. These algorithms
implement in different ways essentially the same idea: The core halo
density is determined by the mean cosmological density at the halo
formation redshift, which is typically defined as the redshift when
the most massive progenitor of the final halo reaches a certain small
fraction of the final halo mass. The factor between the mean
cosmological density and the halo core density, and the fraction of
the final halo mass used for defining the halo formation redshift, are
two free parameters in the algorithms by \cite{NA97.1} and
\cite{BU01.1}. The algorithm by \cite{EK01.1} has only one free
parameter. We determine these parameters by minimising the squared
deviation between the analytic halo concentrations expected for the
given halo masses and redshifts, and the numerically-determined halo
concentrations.

\begin{figure}[ht]
  \centerline
    {\includegraphics[width=\hsize]{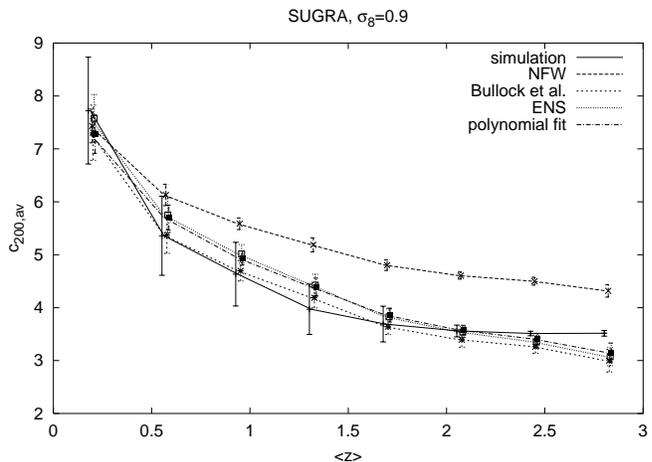}}
\caption{Example for the agreement between analytically expected and
  numerically determined halo concentrations in the SUGRA model
  normalised to $\sigma_8=0.9$. The solid line with the error bars
  shows our numerical results in eight redshift bins between redshifts
  3 and 0. ``NFW'' and ``ENS'' stand for \cite{NA97.1} and
  \cite{BU01.1}, respectively.}
\label{fig:1}
\end{figure}

We find that excellent agreement between the numerical halo
concentrations and the predictions by the algorithm of \cite{BU01.1}
can be achieved if haloes are assumed to form very early, or, in other
words, if we assume that halo core properties are imprinted when only
a very small fraction of the final halo mass is already in place. The
results from the algorithm of \cite{EK01.1} agree very well with the
numerical concentrations without any parameter adaptation, while the
\cite{NA97.1} algorithm predicts too shallow redshift evolution
(cf.~Fig.~\ref{fig:1}).

Our results admit a power-law fit of the form
\begin{equation}
  \bar c(M,z)=\frac{c_0}{1+z}
  \left(\frac{M}{10^{14}\,h^{-1}\,M_\odot}\right)^\alpha\;,
\label{eq:04}
\end{equation}
where $c_0$ is a constant for each cosmology. The exponent
$\alpha\approx-0.1$ for all cosmologies tested, thus the
mass-dependence of the halo concentrations is quite shallow for
massive haloes. Table~\ref{tab:1} summarises the parameters we find.

\begin{table}[ht]
\caption{Parameters $c_0$ and $\alpha$ for the fit formula
  (\ref{eq:04}) for halo concentrations in our eight cosmological
  models.}
\label{tab:1}
\begin{center}
\begin{tabular}{|l|rrr|}
\hline
model & $\sigma_8$ & $c_0$ & $\alpha$ \\
\hline
$\Lambda$CDM  & 0.90 & $ 9.59\pm0.07$ & $-0.102\pm0.004$ \\
Ratra-Peebles & 0.90 & $10.20\pm0.07$ & $-0.094\pm0.005$ \\
Ratra-Peebles & 0.82 & $ 9.30\pm0.06$ & $-0.108\pm0.005$ \\
SUGRA  & 0.90 & $11.15\pm0.09$ & $-0.094\pm0.006$ \\
SUGRA  & 0.76 & $ 9.46\pm0.07$ & $-0.099\pm0.005$ \\
$w=-0.6$ & 0.90 & $11.32\pm0.09$ & $-0.092\pm0.005$ \\
$w=-0.6$ & 0.86 & $10.44\pm0.08$ & $-0.066\pm0.005$ \\
\hline
\end{tabular}
\end{center}
\end{table}

\begin{figure}[ht]
  \centerline
    {\includegraphics[width=\hsize]{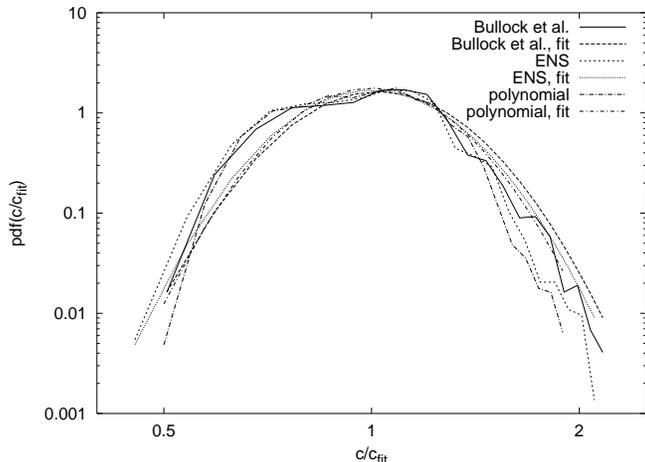}}
\caption{Quite independent of cosmology, the halo concentrations fall
  on a log-normal distribution with a scatter of $\approx0.22$.}
\label{fig:2}
\end{figure}

The scatter among the concentrations is large, but well-described by a
log-normal about the mean given by (\ref{eq:04}). Quite independent of
the cosmological model, the standard deviation of $\ln(c/\bar c)$ is
$\approx0.22$ (cf.~Fig.~\ref{fig:2}), see also \cite{JI00.1}.

Interestingly, it turns out that a simple description can also be
given for the cosmology-dependence of the parameter $c_0$. We find
$c^{\Lambda\mathrm{CDM}}_0\approx9.6$, and for the other cosmological
models
\begin{equation}
  c_0=c^{\Lambda\mathrm{CDM}}_0\,
  \frac{D_+(z_\mathrm{coll})}
  {D^{\Lambda\mathrm{CDM}}_+(z_\mathrm{coll})}\;,
\label{eq:05}
\end{equation}
i.e.~the concentrations scale in proportion to the linear growth
factor, provided that the halo collapse redshift $z_\mathrm{coll}$ is
chosen high enough; in fact, even $z_\mathrm{coll}\to\infty$ produces
good agreement with the numerical simulations. This reinforces the
impression that halo properties are imprinted at very early times,
when only a very small fraction of the final halo mass is already in
place.

The higher halo concentrations found in dark-energy cosmologies are
expected to have pronounced consequences for strong gravitational
lensing by galaxy clusters. This is indicated by an analytic study we
have carried out, and demonstrated by a numerical study using the same
cluster sample described here (cf.~the contribution of Meneghetti et
al.~to these proceedings).

\section{Conclusions}

Using numerical simulations of 17 cluster-sized haloes in eight
different cosmologies, we have tested and confirmed analytic
expectations for the dependence of halo concentrations on models for
the dark energy. Earlier structure growth in dark-energy models yields
more concentrated haloes. If it is assumed that core halo properties
are imprinted at very early times, when only a very small fraction of
the final halo mass is already in place, the halo-concentration
algorithms proposed by Bullock et al. and Eke et al. turn out to work
remarkably well. Halo concentrations have a log-normal distribution
with a scatter of $\approx0.22$ about their mean values, quite
independent of cosmology. For all cosmologies tested, the mean halo
concentration at fixed mass and redshift scales in proportion to the
linear growth factor at the halo formation time if the latter is
defined to be very early.

\section*{Acknowledgements}

We are grateful to Volker Springel for his code and support, and to
Simon White for his constructive comments. The simulations were
carried out on the IBM-SP4 machine at the ``Centro Interuniversitario
del Nord-Est per il Calcolo Elettronico'' (CINECA, Bologna), with CPU
time assigned under an INAF-CINECA grant. K.~Dolag acknowledges
support by a Marie Curie Fellowship of the European Community program
``Human Potential´´ under contract number MCFI-2001-01227.

\bibliography{../TeXMacro/master}

\begin{thebibliography}{14}
\expandafter\ifx\csname natexlab\endcsname\relax\def\natexlab#1{#1}\fi
\expandafter\ifx\csname url\endcsname\relax
  \def\url#1{\texttt{#1}}\fi
\expandafter\ifx\csname urlprefix\endcsname\relax\def\urlprefix{URL }\fi

\bibitem[{Bartelmann et~al.(2002)Bartelmann, Perrotta, and
  Baccigalupi}]{BA02.1}
Bartelmann, M., Perrotta, F., Baccigalupi, C., 2002. Halo concentrations and
  weak-lensing number counts in dark energy cosmologies. A\&A 396, 21.

\bibitem[{Brax and Martin(2000)}]{BR00.2}
Brax, P., Martin, J., 2000. Robustness of quintessence. PRD 61, 103502.

\bibitem[{Bullock et~al.(2001)Bullock, Kolatt, Sigad, Somerville, Kravtsov,
  Klypin, Primack, and Dekel}]{BU01.1}
Bullock, J., Kolatt, T., Sigad, Y., Somerville, R., Kravtsov, A., Klypin, A.,
  Primack, J., Dekel, A., 2001. MNRAS 321, 559.

\bibitem[{Dolag et~al.(2004)Dolag, Bartelmann, Perrotta, Baccigalupi,
  et~al.}]{DO03.2}
Dolag, K., Bartelmann, M., Perrotta, F., Baccigalupi, C., et~al., 2004.
  Numerical study of halo concentrations in dark-energy cosmologies. A\&A 416,
  853.

\bibitem[{Eke et~al.(2001)Eke, Navarro, and Steinmetz}]{EK01.1}
Eke, V., Navarro, J., Steinmetz, M., 2001. ApJ 554, 114.

\bibitem[{Jing(2000)}]{JI00.1}
Jing, Y., 2000. The density profile of equilibrium and nonequilibrium dark
  matter halos. ApJ 535, 30.

\bibitem[{Klypin et~al.(2003)Klypin, Macci{\`o}, Mainini, and
  Bonometto}]{KL03.1}
Klypin, A., Macci{\`o}, A., Mainini, R., Bonometto, S., 2003. Halo properties
  in models with dynamical dark energy. ApJ submitted; preprint
  astro-ph/0303304.

\bibitem[{Navarro et~al.(1997)Navarro, Frenk, and White}]{NA97.1}
Navarro, J., Frenk, C., White, S., 1997. A universal density profile from
  hierarchical clustering. ApJ 490, 493.

\bibitem[{Peebles and Ratra(2002)}]{PE02.2}
Peebles, P., Ratra, B., 2002. The cosmological constant and dark energy.
  Rev.~Mod.~Phys. 75, 599.

\bibitem[{Ratra and Peebles(1988)}]{RA88.1}
Ratra, B., Peebles, P., 1988. Cosmological consequences of a rolling
  homogeneous scalar field. PRD 37, 3406.

\bibitem[{Springel et~al.(2001)Springel, Yoshida, and White}]{SP01.1}
Springel, V., Yoshida, N., White, S., 2001. Gadget: a code for collisionless
  and gasdynamical cosmological simulations. New Astronomy 6, 79.

\bibitem[{Tormen et~al.(1997)Tormen, Bouchet, and White}]{TO97.2}
Tormen, G., Bouchet, F., White, S., 1997. The structure and dynamical evolution
  of dark matter haloes. MNRAS 286, 865.

\bibitem[{Weinberg and Kamionkowski(2003)}]{WE03.1}
Weinberg, N., Kamionkowski, M., 2003. Constraining dark energy from the
  abundance of weak gravitational lenses. MNRAS 341, 251.

\bibitem[{Wetterich(1988)}]{WE88.1}
Wetterich, C., 1988. Cosmology and the fate of the dilatation symmetry. Nuclear
  Physics B 302, 668.

\end{thebibliography}
\bibliographystyle{elsart-harv}

\end{document}